\documentstyle[11pt]{article}
\newdimen\paperwidth
\newdimen\paperlength
\newdimen\margin
 \newdimen\vmargin
\textwidth=\paperwidth
\advance \textwidth by -2\margin
\advance\hoffset by -1truein
\advance\hoffset by \margin \textheight=\paperlength
\advance \textheight by -2\vmargin
\advance\voffset by -1truein
\advance\voffset by \vmargin
\advance\textheight by
-2\baselineskip
\begin{document}

\begin{titlepage} \title{ {\bf Real Space Renormalization Group }\\ {\bf
Methods and Quantum Groups}\thanks{Work partly supported by CICYT under
contracts
AEN93-0776 (M.A.M.-D.) and PB92-109 , European Community Grant
ERBCHRXCT920069
(G.S.).} }

\vspace{2cm} \author{ {\bf Miguel A. Mart\'{\i}n-Delgado}\dag \mbox{$\:$}
and
{\bf Germ\'an Sierra}\ddag \\ \mbox{}    \\ \dag{\em Departamento de
F\'{\i}sica
Te\'orica I}\\ {\em Universidad Complutense.  28040-Madrid, Spain }\\
\ddag{\em
Instituto de Matem\'aticas y F\'{\i}sica Fundamental. C.S.I.C.}\\ {\em
Serrano
123, 28006-Madrid, Spain } } \vspace{5cm}
\date{} \maketitle
\def\baselinestretch{1.3} \begin{abstract}

We apply real-space RG methods to study two quantum group invariant
Hamiltonians,
that of the XXZ model and the Ising model in a transverse field defined
in an
open chain with appropiate boundary terms. The quantum group symmetry is
preserved under the RG transformation except for the appearence of a
quantum
group anomalous term which vanishes in the classical case. We obtain
correctly
the line of critical XXZ models. In the ITF model the RG-flow coincides
with the
tensor product decomposition  of cyclic irreps. of $SU_q(2)$ with
$q^4=1$.

\ \

\ \

\ \

\ \

\ \

\end{abstract}

\vspace{2cm} PACS numbers: 75.10 Jm, 05.50.+q, 64.60.Ak

\vskip-17.0cm \rightline{UCM/CSIC-95-07} \rightline{{\bf July 1995}}
\vskip3in
\end{titlepage}

\newpage
\def\baselinestretch{1.5} \noindent
Real Space Renormalization Group methods, as applied to quantum many-body
Hamiltonians, originated from the successful study of the Kondo problem
by Wilson
\cite{wilson}. Later on people working in field theory and condensed
matter
generalized it to other problems by using the Kadanoff's concept of block
\cite{drell}, \cite{jullien}. The Block method (BRG) has the advantage of
being
conceptually and technically simple, but it lacks of numerical accuracy
or may
even produce wrong results. For this reason the analytical BRG methods
were
largely abandoned in the 80's in favor of numerical methods such as the
Quantum
Monte Carlo approaches. In the last few years there has been new
developments in
the numerical RG methods motivated by a better understanding of the
errors
introduced by the splitting of the lattice into disconnected blocks. A
first step
was put forward in \cite{white-noack} where a combination of different
periodic
boundary conditions applied to every block lead to the correct energy
levels of a
simple tight-binding model. This method however has not been generalized
to
models describing interactions. A further step in this direction was
undertaken
by White in \cite{white} were a Density Matrix algorithm (DMRG) is
developed. The
main idea is to take into account the connection of every block with the
rest of
the system when choosing the states which survive the truncation
procedure. The
standard prescription is tho choose the lowest energy states of the block
Hamiltonian. Instead, in the DMRG method one replaces the block
Hamiltonian by a
block density matrix and chooses the eigenstates of this matrix with the
highest
eigenvalues. The density matrix is constructed out of the ground state of
a
superblock which contains the desired block.

In this letter we want to propose another RG method which uses the
concept of
quantum groups. This mathematical notion emerged in the study of
integrable
systems and it has been applied to conformal field theory, invariants of
knots
and manifolds, etc. \cite{drinfeld-jimbo}, \cite{germancesar}. The new
application of quantum groups that we envisage has been partially
motivated by
the aforementioned work of White, Noack and collaborators an it is
probably
related to it. This relation is suggested by the fact that quantum groups
describe symmetries in the presence of non-trivial boundary conditions.
The
typical example to understand this property of quantum groups is given by
the 1D
Heisenberg-Ising model with anisotropic parameter $\Delta $. The
isotropic model
$\Delta = \pm1$ is invariant under the rotation group $SU(2)$, but as
long as
$|\Delta| \neq 1$ this symmetry is broken down to the rotation group
$U(1)$
around the z-axis. One can ``restore" this full rotation symmetry by
adding
appropiate boundary operators to the Hamiltonian of the open chain. The
classical
group $SU(2)$  becomes then the quantum group $SU_q(2)$, where the
quantum
parameter is related to the anisotropy by $\Delta  = \frac{q+q^{-1}}{2}$
\cite{sklyanin}, \cite{pasquier-saleur}. The ``restoration" of a
classical
symmetry into a $q$-symmetry is achieved at the price of deforming the
algebra
and the corresponding addition rule of angular momentum. The $q$-sum
rule, which
is called the comultiplication, becomes non local and violates parity.
The total
raising (lowering) operators acting on the whole chain are a sum of the
raising (
lowering) operators acting at every single site times  a non-local term
involving
all the remaining sites which appear in an asymmetric way: sites located
to the
left or to the right at a given site contribute differently.

These features of $q$-groups made them specially well-suited to implement
a RG
method which takes into account the correlation between neighboring
blocks. Let
us next show how this can be done explicitly in two examples in 1D: the
Heisenberg-Ising model and the Ising model in a transverse field (ITF).

\vspace{30 pt}

{\em Heisenberg-Ising Model (XXZ Model)}. The open spin chain Hamiltonian
is
defined as:

\begin{equation} H_N(q,J) = \sum _{j=1}^{N-1} h_{j,j+1}(q,J)
\label{1} \end{equation}

\begin{equation} h_{j,j+1}(q,J)  = \frac{J}{2} [\sigma _j^x \sigma
_{j+1}^x +
\sigma _j^y \sigma _{j+1}^y + \frac{q + q^{-1}}{2} \sigma _j^z \sigma
_{j+1}^z -
\frac{q - q^{-1}}{2} (\sigma _j^z -  \sigma _{j+1}^z)]
\label{2} \end{equation}

\noindent where $\vec{\sigma}_j$ are standard Pauli matrices acting at
the
$j^{th}$ site of the chain. For the time being $q$ is an arbitrary
complex
parameter. Observe that the successive terms in $\sigma _j^z - \sigma
_{j+1}^z$
in (\ref{2}) when added into the total $H$ only gives boundary operators
proportional to $\sigma_1^z - \sigma_N^z$ plus the standard bulk
Hamiltonian.

Let us now introduce the $q$-group generator of $SU_q(2)$ acting in the
spin
chain of $N$ sites \cite{pasquier-saleur}.

\begin{equation} S^z =  \frac{1}{2} \sum _{j=1}^N \sigma_j^z
\label{3a}
\end{equation}

\begin{equation} S^{\pm } =  \sum _{j=1}^N q^{-\frac{1}{2} (\sigma_1^z +
\ldots +
\sigma_{j-1}^z)} \sigma_j^{\pm }  q^{\frac{1}{2} (\sigma_{j+1}^z + \ldots
+
\sigma_{N}^z)}         \label{3b} \end{equation}

\noindent which satisfy the quantum group algebra,

\begin{equation} [S^{+}, S^{-}] = \frac{q^{2 S^z} - q^{-2 S^z}}{q -
q^{-1}}
                           \label{4} \end{equation}

\noindent The important fact is that not only the whole Hamiltonian
(\ref{1}) but
also the site-site Hamiltonian (\ref{2}) commutes with the generators
(\ref{3a})-(\ref{4}) of $SU_q(2)$:

\begin{equation} [h_{j,j+1}, S^z] =  [h_{j,j+1}, S^{\pm}] = 0  \ \
\forall j
    \label{5} \end{equation}

\noindent Hence the eigenstates of $H_N(q,J)$ can be classified according
to the
representations of $SU_q(2)$ (see \cite{pasquier-saleur} for details.)

To construct a real space RG for the Hamiltonian (\ref{1}) we shall
choose blocks
of 3 sites. This is important in order to get a renormalized Hamiltonian
of the
same form as the original one. The block Hamiltonian involving the first
3 sites
is simply,

\begin{equation} H_B = h_{12} + h_{23}
               \label{6} \end{equation}

\noindent Now we can apply $q$-group representation theory to diagonalize
$H_B$.
There are 3 energy levels corresponding to the $q$-tensor product
decomposition
$1/2\otimes1/2\otimes1/2 = 1/2\oplus1/2\oplus3/2$.

\noindent For $q$ real and positive (i.e. $\Delta \geq 1$ )  or $q$ a
phase
(i.e.$ |\Delta| \leq 1$) the lowest energy level of  (\ref{6}) is doubly
degenerated and  corresponds to one of the spin $1/2$  irreps, which
reads
explicitly

\begin{equation} |\frac{1}{2} \rangle  =  \frac{1}{\sqrt{2 (q + q^{-1} +
1)}}
\left( -q^{-1/2} | \downarrow \uparrow \uparrow \rangle + (q^{1/2} +
q^{-1/2})
|\uparrow \downarrow \uparrow \rangle - q^{1/2} |\uparrow \uparrow
\downarrow
\rangle     \right)                 \label{7a} \end{equation}

\begin{equation} |-\frac{1}{2} \rangle  = \frac{1}{\sqrt{2 (q + q^{-1} +
1)}}
\left( q^{1/2} | \downarrow \downarrow \uparrow \rangle - (q^{1/2} +
q^{-1/2})
|\downarrow \uparrow \downarrow \rangle + q^{1/2} |\uparrow \downarrow
\downarrow
\rangle     \right)                 \label{7b} \end{equation}

\noindent whose energy is $e_B = -\frac{J}{2} (q + q^{-1} + 2)$.

If we take the $q \rightarrow 0^+$ limit in (\ref{1}) and (\ref{2}) we
obtain an
Ising model Hamiltonian ($\Delta \rightarrow \infty$) with a unique
ground state
given by the Neel state $|\downarrow \uparrow \downarrow \uparrow \ldots
\rangle$
(notice that this  uniqueness is due to the boundary term $\frac{q
-q^{-1}}{2}
(\sigma_1^z-\sigma_N^z)$). On the other hand, for a block of 3 sites
there are 4
states of lowest energy ($|\uparrow \downarrow \uparrow \rangle,
|\downarrow
\uparrow \uparrow \rangle, |\downarrow \uparrow \downarrow \rangle
\mbox{and}
|\downarrow \downarrow \uparrow \rangle$), while for a block of 4 sites
there is
again only one ground state given by $|\downarrow \uparrow \downarrow
\uparrow
\rangle$. This means that choosing an odd number of blocks is not
appropiate to
study the Ising limit of (\ref{2}). To do so one should choose an even
number of
sites, but this will not be pursued here. Hence we shall concentrate on
$q$ being
a phase.

The renormalization prescription consits in choosing the states
(\ref{7a})-(\ref{7b}) as the spin up $|\uparrow \rangle'$ and down
$|\downarrow
\rangle'$ states associated to the whole block as if it were a single
site. Using
the standard methods of BRG we obtain the following RG-transformation
laws for
the spin operators $\vec{S}_i$ acting at the sites $i=1$ and $3$:

\begin{equation} (S_i^x)_{RG} = \xi (q) \   \mbox{$S'$}^x    \ \ \  \
i=1,3
\label{8a} \end{equation}

\begin{equation} (S_i^y)_{RG} = \xi (q) \   \mbox{$S'$}^y    \ \ \  \
i=1,3
\label{8b} \end{equation}

\begin{equation} (S_i^z)_{RG} = \xi (q) \   \mbox{$S'$}^z + \eta_i (q) 1'
\ \ \
 \  i=1,3      \label{8c} \end{equation}

\noindent where $\xi (q)$ is a renormalization factor which depends upon
$q$ as:

\begin{equation} \xi (q) = \frac{q + q^{-1} + 2}{2(q + q^{-1} + 1)}
\label{9}
\end{equation}

\noindent and

\begin{equation} \eta_1 = -\eta_3 \equiv \eta(q) =  \frac{q - q^{-1}}{4(q
+
q^{-1} + 1)}      \label{10} \end{equation}

\noindent The multiplicative renormalization factor $\xi(q) $ is common
to all
the spin operators $\vec{S}_i$ as a consequence of the full symmetry
group
$SU_q(2)$. The ``quantized" feature of $SU_q(2)$ is reflected in the
``quantum
group anomaly" term in (\ref{8c})-(\ref{10}), which indeed shows the
deviation
from the classical case ($q=1$). Eqs. (\ref{8a})-(\ref{8c}) are quite
different
from the standard BRG analoge for the Heisenberg-Ising model done in
reference
\cite{rabin}, where the RG-equations for $S^x$ and $S^y$ differ from
those of
$S^z$ (i.e. $\xi ^x = \xi^y \neq \xi^z$.)

\noindent Using eqs. (\ref{8a})-(\ref{8c}) we can get the renormalized
block-block Hamiltonian $h_{3k,3k+1}$. Putting all terms together we
arrive at
the following effective Hamiltonian $H'$ which acts on the chain having
$N/3$
sites,

\begin{equation} H' = H_{N/3} (q',J') + \frac{N}{3} e_B (q,J) +
(\frac{N}{3} - 1)
\  e_{BB} (q,J)             \label{11} \end{equation}

\noindent where

\begin{eqnarray} q' & = & q                                    \label{12}
\\ J'
& = & \xi^2(q) J \\ e_B (q,J) & = & -\frac{J}{2} (2 + q + q^{-1}) \\
e_{BB} (q,J)
& = & J \frac{(q - q^{-1})^2 (3 q + 3 q^{-1} + 4)}{16 (q + q^{-1} + 1)^2}
\end{eqnarray}

\noindent The $e_B$ contribution to the energy comes from the block part
$H_B$
while $e_{BB}$ is a novel contribution coming from the quantum group
anomaly. The
remarkable feature of eqs. (\ref{11}) and (\ref{12}) is that the coupling
constant $q$, or alternatively $\Delta $ {\em does not flow under the
RG-transformation}, while $J^{(m)}$, which is the value of $J$ after $m$
RG-steps, goes to zero in the limit where $m \rightarrow \infty $, which
in turn
implies that the theory is massless. Hence, eqs. (\ref{12}) predicts
correctly a
{\em line of critical models} in the range $|\Delta| \leq 1$. These
models are
described by Conformal Field Theories (CFT) with central extension $c$
less than
one. If we write the quantum parameter as $q=e^{i\pi/(\mu +1)}$ then
$c=1-\frac{6}{\mu (\mu+1)}$ \cite{alcaraz}. The boundary terms in
(\ref{1})-(\ref{2}) are responsible for this fact.

A non-trivial check of the validity of our RG-method can be given in the
case
where $q=e^{i\pi/3}$ for which $c=0$ (percolation limit). The ground
state energy
can be computed exactly from the constant terms (\ref{11}) (assume that
$N=3^m$
and perform $m$-RG steps), and is given by

\begin{equation} E_0 (N,q=e^{i\pi/3})  = - \frac{3}{4} N +   \frac{3}{4}
\label{13} \end{equation}

\noindent This equation coincides with the exact result obtained  in
\cite{alcaraz} using Bethe ansatz. This means that at least in this
particular
case the truncation to the states $|\pm \frac{1}{2}\rangle$ in
(\ref{7a})-(\ref{7b}) of our $q$RG-method {\em does not imply any kind of
approximation}. This is consistent with the fact that in the CFT with
$c=0$ there
is  a unique state, namely, the ground state. What the $q$RG-method does
is to
pick up that piece of the ground state which projects into a given block.

It can be shown that a consistent representation theory of quantum groups
at root
of unity (i.e., $q^{\mu + 1}=-1$) requires the use of truncated tensor
products
of $q$-group irreps. In the case of $q=e^{i\pi/3}$ this truncation
implies:

\begin{equation} \left( 1/2 \otimes 1/2 \otimes 1/2
\right)_{q=e^{i\pi/3}} = 1/2
\label{14} \end{equation}

\noindent which is precisely the truncation performed when restricting
ourselves
to the states (\ref{7a})-(\ref{7b}). In references \cite{ags} the
representation
theory of $q$-groups was put in one-to-one correspondence with that of
Rational
Conformal Field Theories (RCFT). There it was observed that the
truncation
inherent in the construction of the RCFT's has a parallel in the
truncation of
the representation theory of $q$-groups with $q$ a root of unity. The
result we
have obtained in this letter suggests that $q$-group truncations can be
carried
over a RG analysis of $q$-group invariant chains. In other words, using
$q$-groups we can safely truncate states in the block RG method. We may
summarize
this discussion squematically by,

\begin{equation} \mbox{$q$RG-truncation} \leftrightarrow \mbox{RCFT}
                                         \label{15} \end{equation}

\vspace{30 pt}

{\em Ising model in a transverse field (ITF)}. This simple model has been
widely
used to test the validity of BRG methods \cite{drell}, \cite{jullien}.
The
Hamiltonian of an open chain is given by $H=\sum _{j=1}^{N-1} h_{j,j+1} $
where

\begin{equation} h_{j,j+1} = -(J \sigma_j^x \sigma_{j+1}^x + p \sigma
_j^z + p'
\sigma_{j+1}^z )
\label{16}
\end{equation}

\noindent The standard choice is $p=p'=\Gamma/2$, in which case
(\ref{16}) has 4
different eigenvalues.  The BRG method with a block with two sites
chooses just
the 2 lowest ones. However if $(p,p') = (\Gamma,0)$ (or $(0,\Gamma )$)
the
Hamiltonian (\ref{16}) has two doubly degenerate eigenvalues $\pm e_B$
($e_B=\sqrt{J^2+\Gamma^2}$). This choice is not parity invariant but it
implements the self-duality property of the ITF model, yielding the exact
value
of the critical fixed point of the  ITF which appears at $(\Gamma/J)_c=1$
\cite{fpacheco}. This degeneracy of the spectrum of (\ref{16}) has a {\em
q-group
origin}. The relevant quantum group is again $SU_q(2)$ with $q^4=1$.
However the
representations involved are not a $q$-deformation of the spin $1/2$
irrep. as in
the previous example,  but rather a new class of irreps. which only exist
when
$q$ is a root of unity. They are called {\em cyclic irreps.} and neither
are
highest weight nor lowest weight representations as the more familiar
regular
irreps. If we call $E$, $F$ and $K$ the generators of $SU_q(2)$, which
correspond
essentially to $S^{+}$, $S^-$ and $q^{2 S^z}$ in the notation of the
previous
example, then a cyclic irrep. acting at a single site of the chain is
given by:

\begin{equation} E_j = a \sigma_j^x , \ \ F_j = b \sigma_j^y , \ \ K_j =
\lambda
\sigma_j^z       \label{17} \end{equation}

\noindent where $a=\frac{1}{2} \sqrt{\lambda ^2-1}$, $b=-\frac{1}{2}
\sqrt{1-\lambda ^{-2}}$. The parameter $\lambda $ is the label of the
cyclic
irrep. Using (\ref{17}) and the addition rule of $SU_q(2)$ we can get the
representation of $E$, $F$ and $K$ acting on the whole chain:

\begin{eqnarray} E & = & a \sum_{j=1}^N \lambda ^{j-1} \sigma_1^z \cdots
\sigma_{j-1}^z \sigma_j^x   \label{18} \\ F & = & b \sum_{j=1}^N \lambda
^{j-N}
\sigma_j^y \sigma_{j+1}^z \cdots \sigma_{N}^z \\ K & = & \lambda ^N
\prod_{j=1}^N
\sigma_j^z \end{eqnarray}

\noindent Now it is a simple exercise to check that these operators
commute with
(\ref{16}),

\begin{equation} [h_{j,j+1}, E] = [h_{j,j+1}, F] = [h_{j,j+1}, K] = 0, \
\
\forall j                    \label{19} \end{equation}

\noindent assuming that we choose

\begin{equation} \lambda = \Gamma /J                   \label{20}
\end{equation}

\noindent The last of the equalities in (\ref{19}) expresses the
well-known
${\cal Z}_2$-symmetry of the ITF-model which allows one to split the
spectrum of
the Hamiltonian into an even and odd subsectors. {\em The other two
symmetries
are new} and explain the degeneracy of the spectrum of $h_{j,j+1}$. By
all means
the whole Hamiltonian $H=\sum_j h_{j,j+1}$ is also invariant under
(\ref{18}).
Notice that $H$ differs from the standard ITF simply in a term at one of
the ends
of the chain. This is the same mechanism as for the XXZ Hamiltonian: one
needs
properly chosen operators at the boundary in order to achieve quantum
group
invariance. Similarly as for the XXZ model the RG-analysis of the ITF
becomes a
problem in representation of quantum groups: blocking is equivalent to
tensoring
representations. What is the tensor product of cyclic irreps.? Here it is
important to realize that all cyclic irreps. of $SU_q(2)$ have dimension
2, what
distinguishes them is the value of $\lambda $. The tensor product
decomposition
of two cyclic irrep. $\lambda_1$ and $\lambda _2$ is given by:

\begin{equation} [\lambda _1] \otimes [\lambda _2] = 2 \   [\lambda _1
\lambda
_2]               \label{21} \end{equation}

\noindent where the 2 means that $\lambda_1 \lambda_2$ appears twice in
the
tensor product. If we perform a blocking of two sites we will get two
cyclic
irreps. corresponding to $\lambda ^2$. Then we expect from $q$-group
representation theory that the new effective Hamiltonian $h'_{j,j+1}$
will have
the same form as (\ref{16}) but with new renormalized coupling constants
$J'$ and
$\Gamma '$ satisfying:

\begin{equation} \lambda' = \frac{\Gamma '}{J'} = (\frac{\Gamma }{J})^2 =
\lambda^2        \label{22} \end{equation}

\noindent This is indeed the result obtained in \cite{fpacheco}. We
arrive
therefore at the conclusion that {\em the RG-flow of the ITF Hamiltonian
(\ref{16}) is equivalent to the tensor product decomposition of cyclic
irreps of
$SU_q(2)$}. This $q$-group interpretation of the RG-flow is independent
of the
size of the blocks: for a $n$-site block the RG-flow would be $\lambda
\rightarrow \lambda^n$. The fixed point $\lambda =1$ of (\ref{22})
describes the
critical regime of the ITF Hamiltonian and it corresponds to a {\em
singular
point in the manifold of cyclic irreps.}\cite{kac-deconcini},
\cite{kac-deconcini}. At $\lambda =1$ the operators (\ref{18}) are still
symmetries of the Hamiltonian ($a$, $b$ taking any non-zero value) and
they
recall the Jordan-Wigner map between Pauli matrices and 1d-lattice
fermions.

Cyclic irreps. were used in \cite{jimbo} to derive the Boltzmann weights
of the
${\cal Z}_N$-chiral Potts model \cite{macoy}. In \cite{jimbo} the labels
of the
cyclic irreps. have the meaning of rapidities rather than coupling
constants as
in our realization.

The ${\cal Z}_2$ CP-model is nothing but the ITF model. We may wonder
whether the
general ${\cal Z}_N$-chiral Potts model admits a $q$-group RG treatment
along the
lines of this work. This problem will be considered elsewhere. A model
that
admits a qRG analysis is the XY model with a magnetic field $h$ (XYh).
The
results will be presented in \cite{nos}. It suffices to say here that the
$q$-group underlying the model is $SU_q(2)$ with $q^4=1$ and the
representation
used are the so called {\em nilpotent irreps.} \cite{germancesar}, which
are also
described by a parameter $\lambda $ analoge to that in (\ref{17}) and
{\em
related to the magnetic field $h$}. The XYh model is equivalent to a free
fermion
with chemical potential. The results we are obtaining  can be translated
into a
$q$-group symmetry between fermions, either free as in the XY or ITF
models or
interacting as the XXZ model. Another interesting model of interacting
fermions
is the Hubbard model, which has been studied using RG-methods in
\cite{hirsch} .
The integrability of the 1D Hubbard model \cite{liebwu} suggests that it
might be
studied using our qRG techniques.

All the Hamiltonians analysed in this letter are one-dimensional, so they
are
quantum groups as we know them. Despite the fact that the Yang-Baxter
equation
(the precursor of $q$-groups) has a higher dimensional analogue called
the
Zamolodchikov or tetrahedron equation \cite{zamo}, the corresponding high
dimensional analogue of quantum groups is not known. This fact represents
a
barrier to a qRG analysis of Hamiltonians defined in dimensions higher
than one.

\noindent Another possibility, which is suggested by our results, would
be to
define quantum groups as those which contain symmetries which are
anomalous under
RG transformations. This definition is independent of the space
dimensionality.
The quantum anomalous term in equation (\ref{8c}) gives a discrete
realization of
this idea. A continuum analogue of this anomaly is given by the
Feigin-Fuchs
current, which has an anomalous operator product expansion with the
energy-momentum tensor \cite{dotsenko-fateev}. At this point it may be
worth to
recalling  the continous version of quantum groups in CFT of reference
\cite{g-s}, which uses the Feigin-Fuchs or free field realization of the
latter.
Putting all these arguments together, we arrive at the conclusion that
quantum
groups are indeed defined by symmetries anomalous under RG
transformations. This
point of view about quantum groups may set up the pathway to new
developments in
the field.

%(\ref{2})

\vspace{2cm}

{\em e-mail addresses}:  mardel@fis.ucm.es (M.A. M.-D.);
sierra@cc.csic.es (G.S.)

\end{document}